# Complementary Vibrational Spectroscopy

Kazuki Hashimoto[1,2], Venkata Ramaiah Badarla[3], Akira Kawai[1] and Takuro Ideguchi[*3,4]

[1] Department of Physics, The University of Tokyo, Tokyo 113-0033, Japan

[2] Aeronautical Technology Directorate, Japan Aerospace Exploration Agency, Tokyo 181-0015, Japan

[3] Institute for Photon Science and Technology, The University of Tokyo, Tokyo 113-0033, Japan

[4] PRESTO, Japan Science and Technology Agency, Saitama 332-0012, Japan

*ideguchi@gono.phys.s.u-tokyo.ac.jp

**Vibrational spectroscopy, comprised of infrared absorption and Raman scattering spectroscopy, is widely used for label-free optical sensing and imaging in various scientific and industrial fields. The group theory states that the two molecular spectroscopy methods are sensitive to vibrations categorized in different point groups and provide complementary vibrational spectra. Therefore, complete vibrational information cannot be acquired by a single spectroscopic device, which has impeded the full potential of vibrational spectroscopy. Here, we demonstrate simultaneous infrared absorption and Raman scattering spectroscopy that allows us to measure the complete broadband vibrational spectra in the molecular fingerprint region with a single instrument based on an ultrashort pulsed laser. The system is based on dual-modal Fourier-transform spectroscopy enabled by efficient use of nonlinear optical effects. Our proof-of-concept experiment demonstrates rapid, broadband and high spectral resolution measurements of complementary spectra of organic liquids for precise and accurate molecular analysis.**

Vibrational spectroscopy is a fundamental method for chemical analysis used in a variety of scientific fields such as organic/inorganic chemistry, geology, biomedical, material, food, environmental and forensic science[1-5]. The label-free noninvasive molecular spectroscopy enables us to acquire bond specific chemical information of specimen, and it is known that infrared (IR) absorption and Raman scattering spectroscopy provide complementary information of molecular vibrations: the former is active for anti-symmetric vibrations that alter the dipole moment, while the latter for symmetric vibrations that alter the polarizability[1]. IR absorption spectroscopy, which is active for polar bonds such as O-H or N-H, is often used for identification of functional groups of molecules, while Raman scattering spectroscopy, active for bonds such as C=C, S-S or C-S[4], is used for identification of skeletal structures. The group theory states that each fundamental vibrational mode in a molecule with the center of symmetry is either IR or Raman active, and no mode is active for both of them (the rule of mutual exclusion)[1]. Therefore, to acquire the complete information of molecular vibrations for more accurate and precise chemical analysis, both the IR and Raman spectra must be measured. Measuring the complete information of molecular vibrations enables us to analyze complex molecular phenomena such as catalytic chemical reactions[6-9].

Simultaneous measurement of IR and Raman spectra is a grand challenge in spectroscopy because wavelength regions of these two spectroscopy methods are largely separated, i.e., mid-infrared (2.5-25 μm, corresponding to 400-4,000 cm$^{-1}$) for IR spectroscopy and visible to near-infrared (0.4-1 μm, corresponding to 25,000-10,000 cm$^{-1}$) for Raman spectroscopy, respectively. Since this large wavelength discrepancy causes the difficulty of sharing light sources and optics, a primitive combination of conventional FT-IR and Raman spectrometers[10, 11] has never been a convincing approach. Such a system requires a complex instrument comprised of different spectroscopy methods based on a Michelson interferometer and a dispersive spectrometer with two independent light sources such as an incoherent lamp source and a visible continuous-wave laser. Additionally, these conventional methods do not provide state-of-the-art sensitivity and data acquisition speed because of the low brightness of the lamp source for FT-IR and the inherent weakness of spontaneous Raman scattering. Meanwhile, the technical advancement of nonlinear optics based on ultrashort pulsed lasers has enabled us to have higher brightness of coherent IR sources and stronger Raman signals through the coherent Raman scattering[5, 12], and some approaches have been made towards IR/Raman dual-modal spectral acquisition with a single pulsed laser[13, 14]. However, these techniques do not have capability of simultaneous acquisition of complementary IR/Raman spectra nor broadband spectral acquisition covering the molecular fingerprint region (800-1,800 cm$^{-1}$), where the richest vibrational modes exist.

Here, we propose and demonstrate a simple yet powerful technique, called complementary vibrational spectroscopy (CVS), that allows us to simultaneously measure broadband IR and Raman spectra covering the fingerprint region at the same position. CVS is dual-modal Fourier-transform spectroscopy (FTS) enabled by an ultrashort near-infrared (NIR) pulsed laser and a Michelson interferometer. The IR spectroscopy is implemented as FT-IR with a bright mid-infrared (MIR) pulsed source generated via intra-pulse difference-frequency generation (IDFG) from the NIR pulses[15-17], while the Raman spectroscopy as Fourier-transform coherent anti-Stokes Raman scattering spectroscopy (FT-CARS) with the same NIR pulses[18]. The former uses a second-order and the latter a third-order nonlinear phenomena, respectively. The system is simple and robust because it shares a single laser source and an interferometer. Note that our proposed method can be applied to advanced FTS techniques such as dual-comb spectroscopy[19-21], empowering the emerging techniques further.

The schematic representation of the system is shown in Figure 1**a**. In CVS, FT-IR and FT-CARS setups are both implemented in a single FTS system that consists of a Michelson interferometer with a NIR femtosecond laser (10-fs Ti:Sapphire mode-locked laser at a repetition rate of 75 MHz in this study) as a light source. The pulses emitted from the laser are coupled in the interferometer and double NIR pulses are created from each pulse with an optical path length difference (OPD) set by the delay line in the interferometer. The NIR double pulses are focused onto a $\chi^{(2)}$ nonlinear crystal (GaSe crystal in this case) and a portion of the NIR pulses are converted to MIR pulses through the IDFG process. The generated MIR and undepleted NIR pulses collinearly irradiate the sample. The MIR pulses are absorbed, while the NIR pulses are inelastically (Raman) scattered by molecules. The MIR and NIR pulses are spatially separated by a dichroic mirror after passing through the sample and are simultaneously detected by a

HgCdTe (MCT) photodetector and a Si avalanche photodetector (APD), respectively. Here, the NIR pulses are optically filtered before the detector so that only the blue-shifted scattered photons reach the APD. The detected signals are A/D converted by a digitizer and the digitized interferograms are Fourier-transformed. The full schematic of the CVS is depicted in Supplementary Figure 1.

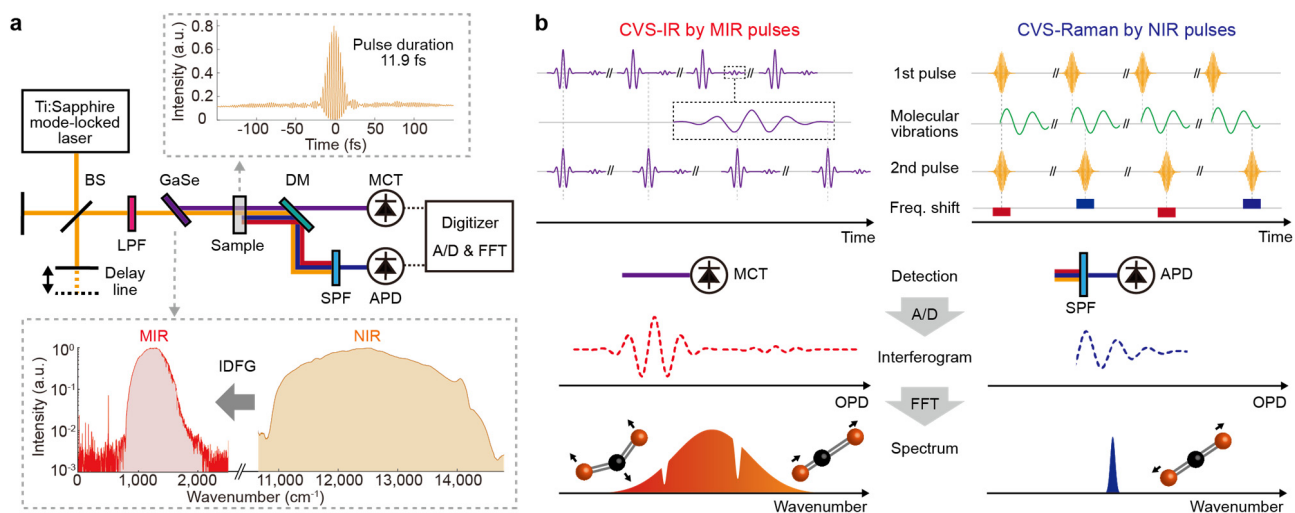

**Figure 1| Schematic and concept of CVS. a,** Schematic of CVS. The insets show the autocorrelation trace of the NIR pulses and the spectra of NIR and MIR pulses. BS: Beamsplitter, LPF: Long-pass filter, DM: Dichroic mirror, SPF: Short-pass filter, MCT: HgCdTe photodetector, APD: Avalanche photodetector. **b,** Conceptual description of CVS. The figure shows a linear triatomic molecule as an example of molecular vibrations. The left panel displays CVS-IR process given by MIR pulses, while the right panel CVS-Raman process by NIR pulses. OPD: Optical path length difference.

The working principle of CVS is shown in Figure 1**b**. In CVS-IR process, the MIR double pulses are modulated by IR-active molecular absorptions and their optical interference is detected by the MIR detector. Since the delay between the first and second MIR pulses is determined by that of the NIR pulses, the MIR absorption interferogram is measured as a function of the OPD between the NIR double pulses. Fourier-transforming the IR interferogram shows a broadband IR spectrum. On the other hand, in CVS-Raman process, the first NIR pulse excites the molecular vibrations and the second NIR pulse probes them and generates blue-shifted photons via anti-Stokes Raman scattering. By scanning the OPD between the NIR pulses, optical frequency of the second NIR pulse can be shifted by the refractive index modulations caused by the Raman active molecular vibrations induced by the impulsive stimulated Raman scattering process. A CARS interferogram that represents the molecular vibrations is obtained as an intensity modulation of the blue-shifted part of the second NIR pulses, which can be separated out by the optical short-pass filter. Finally, a broadband Raman spectrum is obtained by Fourier-transforming the CARS interferogram.

We first characterize the NIR and MIR pulses. The spectrum of our 10-fs Ti:Sapphire laser spans over 10,870 – 14,490 $cm^{-1}$ (690 – 920 nm) at the center wavelength of 12,500 $cm^{-1}$ (800 nm), and its pulse duration is evaluated by autocorrelation measurement as 11.9 fs at the sample position. This ultrashort NIR pulses with the broadband spectrum spanning more than 3,400 cm$^{-1}$ allows us to measure broadband FT-CARS spectrum covering the molecular fingerprint region (800-1,800 cm$^{-1}$) and C-H stretching region (2,800-3,300 cm$^{-1}$)[22]. The spectrum of the MIR pulses generated by the IDFG process in a 30-μm GaSe crystal is measured by a homemade FT-IR spectrometer and it spans from 790 to 1,800 cm$^{-1}$, which covers the fingerprint region. In this study, the lowest wavenumber of the IR spectrum is limited by the detection range of the MCT detector and the highest wavenumber is possibly limited by the phase-matching condition of the IDFG process. Note that the IR spectral region can be shifted by changing the angle of the crystal and also expanded by changing the crystal and/or laser system.

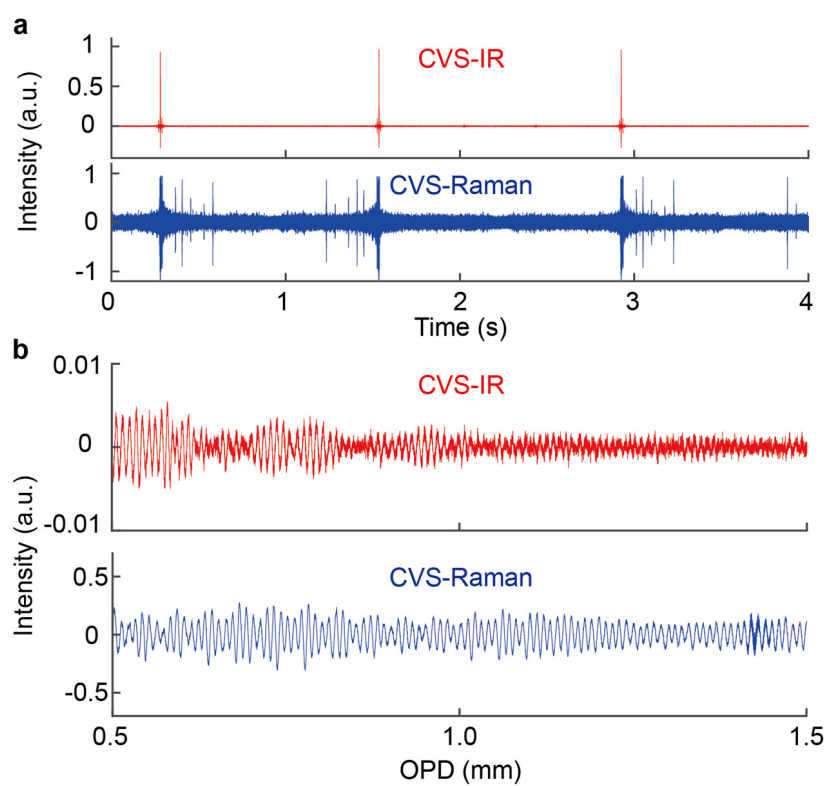

**Figure 2| CVS interferograms of toluene. a**, The upper and lower panels show the sequential interferograms measured by CVS-IR and CVS-Raman spectroscopy, respectively. **b,** Zoomed 15-averaged CVS-IR and CVS-Raman interferograms plotted as a function of the OPD.

As a proof of concept demonstration, we measure complementary vibrational spectra of liquid toluene. Figure 2**a** shows sequential CVS interferograms of toluene, where the IR and CARS interferograms are simultaneously detected. They show the synchronized bursts at zero-path-difference (ZPD) of the interferometer. The OPD is scanned over 2 cm at a rate of 0.8 Hz. Figure 2**b** shows 15-times coherently averaged IR/CARS interferograms, which clearly show signature of molecular vibrations. In the configuration where the MIR pulses are generated after the NIR interferometer, the raw IR temporal waveform contains other components than the desired IR interferogram. A detailed retrieval procedure of the IR interferogram is described in Supplementary Note 4 and Supplementary Figure

2. The double-sided IR interferogram is apodized and Fourier-transformed, whereas the single-sided CARS interferogram is apodized and Fourier-transformed by omitting the center-burst caused by the non-resonant four-wave mixing process at ZPD.

The complementary vibrational spectra Fourier-transformed from the interferograms shown in Figure 2 are displayed in Figure 3 with the reference spectra individually measured by conventional spectrometers. The upper panel shows the CVS-IR spectrum together with the reference spectrum measured by a standard FT-IR spectrometer (FT/IR-6800, JASCO). The CVS-IR transmittance spectrum of toluene agrees well with the reference spectrum and clearly displays the vibrational modes of, e.g., C-H bending at 1,082 and 1,179 cm$^{-1}$ and ring stretching at 1,495 and 1,605 cm$^{-1}$ [23-26] with a triangular-apodized spectral resolution of 3.5 cm$^{-1}$. The lower panel shows the CVS-Raman spectrum and the reference spectrum measured by a standard spontaneous Raman spectrometer (inVia, Renishaw). The CVS-Raman spectrum at the apodized spectral resolution of 5.5 cm$^{-1}$ clearly shows the vibrational modes of ring stretching at 1,003 cm$^{-1}$ and 1,030 cm$^{-1}$, C-CH$_3$ stretching at 1,211 cm$^{-1}$ and CH$_3$ deformation at 1,380 cm$^{-1}$ [23,25,26], which also agrees well with the reference spectrum. Note that the wavenumbers of the CVS spectra are calibrated with the same interferometer, so that we can compare the spectra in a precise manner.

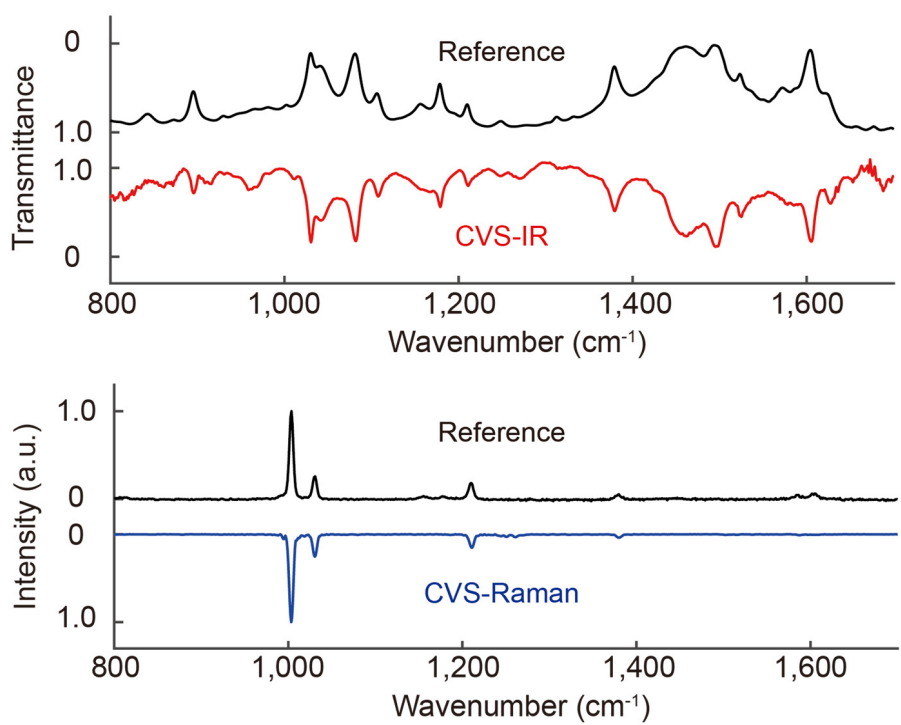

**Figure 3| Complementary vibrational spectra of toluene in the fingerprint region.** The upper panel shows the comparison of the CVS-IR spectrum and the reference IR absorption spectrum measured by a standard FT-IR. The lower panel shows the comparison of the CVS-Raman spectrum and the reference Raman scattering spectrum measured by a spontaneous Raman spectrometer.

To show the applicability of this system for other samples, we measure the complementary spectra of three different kinds of liquid samples: benzene, chloroform and a 4:1 mixture of benzene and DMSO (Figure 4). Here we show the CVS-Raman spectra up to around 3,000 cm$^{-1}$, showing its ultra-broadband measurement capability. This capability of measuring ultra-broadband spectra with high spectral resolution is a unique feature given by using Fourier-transform spectroscopy technique. Detailed assignment of these spectra is discussed in Supplementary Note 2.

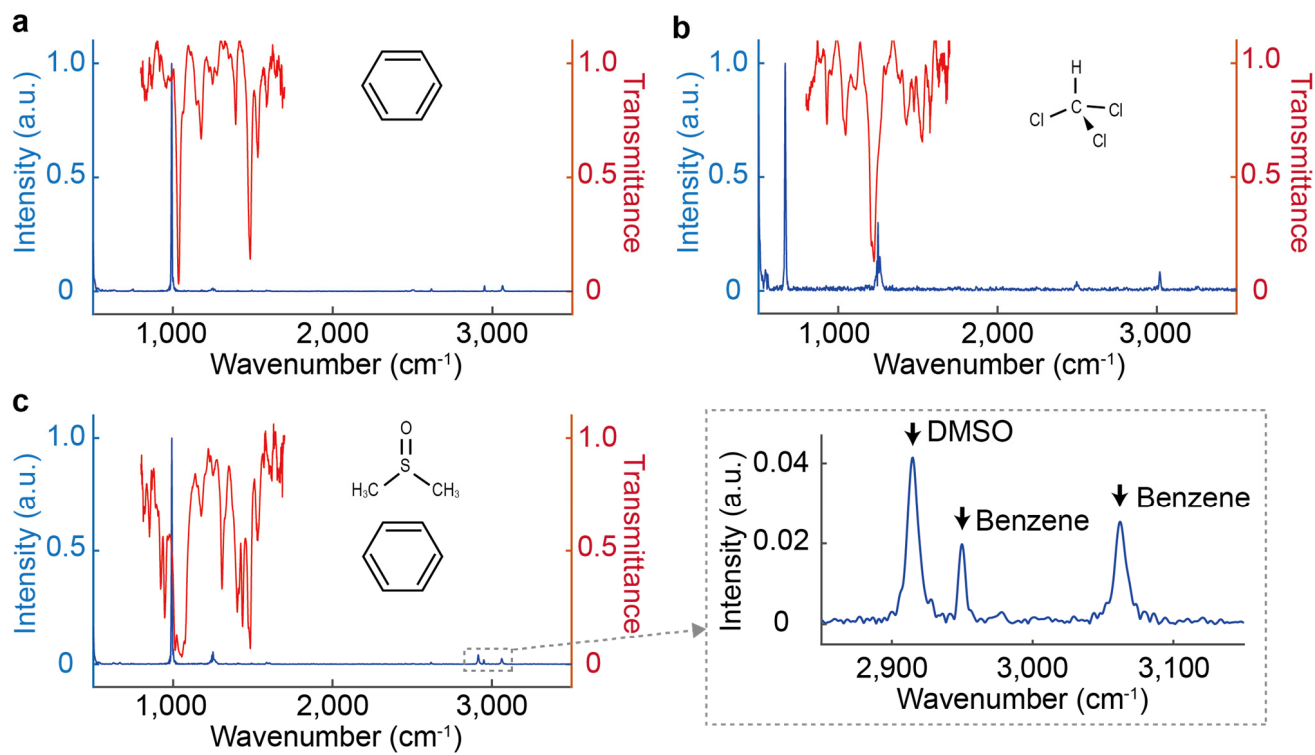

**Figure 4| Complementary vibrational spectra of organic molecules. a,** benzene, **b,** chloroform, **c,** 4:1 mixture of benzene and DMSO. The inset shows the zoomed spectrum of the Raman peaks of DMSO and benzene around 3,000 cm$^{-1}$. The vibrational lines are found from 800 to 1,700 cm$^{-1}$ in the CVS-IR spectrum and from 600 to 3,100 cm$^{-1}$ in the CVS-Raman spectrum. Small peaks at 1,250 cm$^{-1}$ and 2,500 cm$^{-1}$ are attributed to instrumental noises, which can be removed by careful instrumentation.

Our CVS technique can be improved further. The measurable spectral span of the IR absorption spectrum (limited to 1,000 cm$^{-1}$ in this study) can be largely expanded by using other nonlinear crystals or light sources[27-31], and/or by implementing the EO-sampling technique for detecting the IR interferograms[17,27,29]. In addition, since the CVS is based on FTS, it can be more rapid, robust and compact by using dual-comb FTS[19-21] or ultra-rapid-scan FTS[22,32-34]. The higher scan rate allows us to measure dynamically changing complex phenomena or hyper-spectral multi-modal wide-area images. Furthermore, we could use the second harmonic pulses generated from the nonlinear crystal together with the IDFG pulses for adding other spectroscopic modalities in CVS.

**Methods**

**NIR interferometer**

A detailed schematic of the CVS is shown in Supplementary Figure1. The system is based on dual-modal FTS equipped with an ultrashort NIR pulsed laser, a Michelson-interferometer, a nonlinear crystal for MIR pulse generation and photodetectors. The NIR pulsed laser (Ti:Sapphire Kerr-lens mode-locked laser) generates ultrashort pulses centered at 800 nm with a pulse duration of 10 fs at a repetition rate of 75 MHz (Synergy Pro, Spectra-Physics). The NIR pulses are coupled into the Michelson interferometer to generate double NIR pulses with an OPD between the first and second pulses, which is scanned by a motorized stage in the interferometer. The OPD is measured at interferometric precision by continuous-wave (CW) interferograms of a HeNe laser that monitor the motion of the scan mirror. The measured OPD is used for phase error correction of the interferograms. By using a polarization beamsplitter and quarter-wave plates to construct the Michelson interferometer, the double pulses generated from the Michelson interferometer are orthogonally polarized to each other. A long-pass filter at a cutoff wavelength of 700 nm (FELH0700, Thorlabs) and a chirped mirror pair (DCMP175, Thorlabs) tailors the NIR pulses spectrally and temporally. After the long-pass filter the spectral range of the NIR pulses spans from 10,870 to 14,340 cm$^{-1}$.

**MIR pulse generation**

The NIR double pulses are focused onto the 30-μm GaSe crystal (EKSMA OPTICS) using an off-axis parabolic mirror (OAPM) with a focal length of 25.4 mm for generating MIR pulses. The pulse energy of the NIR pulse is 2-5 nJ, and that of the generated MIR pulse is about 50 fJ. The remaining NIR pulses and generated MIR pulses are collimated by another OAPM with a focal length of 25.4 mm. Since the NIR pulses are slightly chirped by passing through the GaSe crystal, it is compensated by another chirped mirror pairs by separating the NIR and MIR pulses with a NIR/MIR dichroic mirror. The NIR and MIR pulses are combined again by another dichroic mirror. Here, to avoid spurious nonlinear effects at the sample, we intentionally have the NIR and MIR pulses separated in time. A pulse duration of the NIR pulses at the sample position is 11.9 fs evaluated by fringe-resolved autocorrelation measurement.

**Irradiation onto the sample**

The NIR and MIR pulses are focused onto the sample by an OAPM with a focal length of 15 mm. The NIR pulse energies irradiated onto the sample are 1.1 nJ and 2.4 nJ for the first and second pulses, respectively. The liquid samples are contained in a cuvette made of 3-mm thick KBr windows, which is transparent in a wide spectral range covering both NIR and MIR. A teflon spacer with a thickness of 50 μm is inserted between the KBr windows for adjusting the sample thickness.

**Acquisition of the interferograms**

The focused light onto the sample is collected and collimated by another OAPM with a focal length of 15 mm. After the collimation, the NIR and MIR pulses are spatially separated by a NIR/MIR dichroic mirror. The MIR pulses are detected by a $N_2$-cooled MCT detector (KLD-0.5-J1-3/11, Kolmar Technologies), while the optically filtered NIR pulses are detected by an APD (APD410A2/M, Thorlabs), respectively. To detect the anti-Stokes scattering only, a short-pass filter (FESH0700, Thorlabs) is inserted in front of the APD. The detectors' signals are low-pass-filtered and A/D-converted by a digitizer (ATS9440, AlazarTech). The CW interferogram of the HeNe laser is also digitized simultaneously. A part of the system where the MIR pulses travel through is enclosed by a box and purged with $N_2$ gas in order to suppress undesired absorptions of the ambient gases, especially $H_2O$ vapor.

**Data processing**

The sequentially measured interferograms are segmented into single interferograms and coherently averaged. The reference CW interferogram of HeNe laser is used for resampling the digitized interferograms. The MIR interferogram is processed as double-sided, while the CARS interferogram as single-sided. The MIR interferogram is retrieved numerically (See Supplementary Note 4 and Supplementary Figure 2). A strong peak that appears at ZPD in the CARS interferogram caused by the non-resonant background is omitted from the Fourier-transform window. Both the interferograms are apodized by triangular function and Fourier-transformed. To show the IR transmittance, we measure spectra with and without the sample.

**Acknowledgement**

We thank Makoto Kuwata-Gonokami, Junji Yumoto, Kuniaki Konishi and Yusuke Morita for allowing us to use their equipment and giving helpful suggestions. V. R. B. thanks the financial support by JSPS. This work was financially supported by JST PRESTO (JPMJPR17G2), JSPS KAKENHI (17H04852, 17K19071, 17F17028).

**Author contributions**

T.I. conceived the idea of this work. K.H. and T.I. designed the CVS system. V.R.B. calculated the IDFG generation, built the $N_2$-purge system and the crystal mount. A.K. measured the reference IR/Raman spectra and made the sample cuvette. K.H. conducted the CVS experiment and analyzed the data. T.I. supervised the work. K.H. and T.I. wrote the manuscript and all authors contributed to improvement of the manuscript.

**Competing financial interests**

Authors declare no competing interests.

# Supplementary Information

**Supplementary Note 1: Detailed schematic of CVS**

The full schematic of CVS is depicted in Supplementary Figure 1. The detail of the system is described in Methods section.

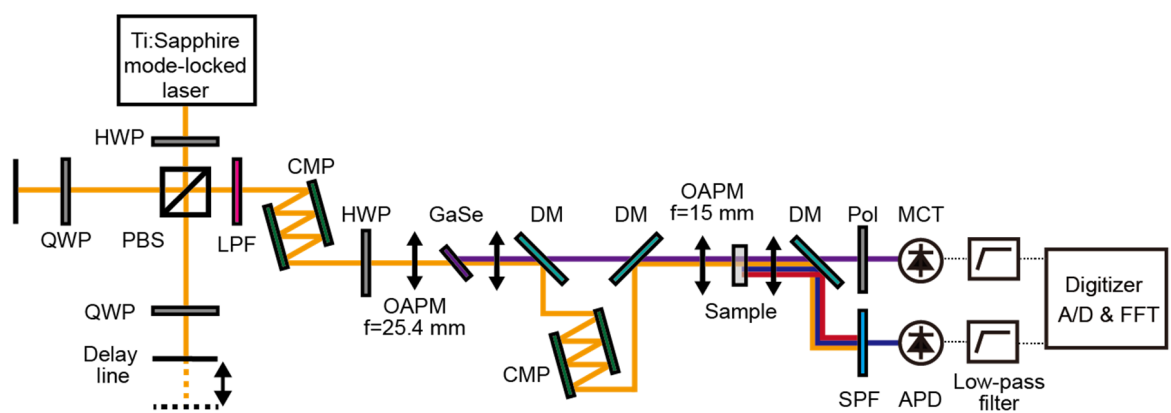

**Supplementary Figure 1| Detailed schematic of CVS.** HWP: Half-wave plate, PBS: Polarizing beamsplitter, QWP: Quarter-wave plate, LPF: Long-pass filter, CMP: Chirped-mirror pair, OAPM: Off-axis parabolic mirror, DM: Dichroic mirror, Pol: Polarizer, SPF: Short-pass filter, MCT: HgCdTe, APD: Avalanche photodetector.

**Supplementary Note 2: Spectroscopic line assignment of the CVS spectra**

We assign some major lines of the measured CVS spectra shown in Figure 4. There are vibrational modes in the CVS-Raman spectrum of benzene at 992 cm$^{-1}$ (ring stretching), 3,063 cm$^{-1}$ (C-H stretching)[1], and 2,950 cm$^{-1}$ (combination or overtone band)[2], whereas in the CVS-IR spectrum at 1,036 cm$^{-1}$ (C-H bending), 1,484 cm$^{-1}$ (ring stretching+deformation)[1]. For chloroform, Raman lines at 667 cm$^{-1}$ (C-Cl stretching) and 3,019 cm$^{-1}$ (C-H stretching), and IR lines at 1,224 cm$^{-1}$ (C-H bending) are observed[1]. For the 4:1 mixed sample of benzene and DMSO, the peaks derived from DMSO are observed at 666 cm$^{-1}$ (C-S stretching), 2,915 cm$^{-1}$ (C-H stretching) in the Raman spectrum[3], and at 947 and 1,054 cm$^{-1}$ (S-O stretching + CH$_3$ rocking), at 1,307 cm$^{-1}$ (CH$_3$ umbrella) and 1,403 and 1,436 cm$^{-1}$ (CH$_3$ bending) in the IR spectrum[4] in addition to the specific peaks of benzene. The peaks that appear at 1,250 cm$^{-1}$ and 2,500 cm$^{-1}$ are attributed to the instrumental noises.

**Supplementary Note 3: Comparison of multi-modal vibrational spectroscopy**

Supplementary Table 1 shows comparison of the spectroscopic specifications of our CVS against the previously demonstrated multi-modal vibrational spectroscopy. As clearly seen in the table, our CVS has broader spectral bandwidth spanning over 1,000 cm$^{-1}$ with higher spectral resolution of around a few cm$^{-1}$, leading to the larger number of spectral elements than those of the previously demonstrated techniques.

Supplementary Table 1 | Spectroscopic specifications of CVS and other multi-modal vibrational spectroscopy techniques.

| Methods | IR spectroscopy / Raman spectroscopy | Observable spectral region (cm$^{-1}$) | Spectral bandwidth (cm$^{-1}$) | Spectral resolution (cm$^{-1}$) | # of spectral elements |
|---|---|---|---|---|---|
| **CVS** (This work) | FT-IR | 800-1,800 | 1,000 | 2 (Unapodized) | 500 |
| | FT-CARS | 600-3,100*$^1$ | 2,500*$^1$ | 3.1 (Unapodized) | 800 |
| [Ref.5] | TSFG | 2,750-3,000 | ~250*$^2$ | ~20 | ~13 |
| | CARS | 2,750-3,000 | ~250*$^2$ | ~20 | ~13 |
| [Ref.6] | Spectral focusing IR | 1,300-3,500 | ~2,200 | ~25 | ~88 |
| | Spectral focusing CARS | 2,250-3,250 | ~1,000 | ~25 | ~40 |

*1: The lowest and highest wavenumber of the observable spectral region of CVS is evaluated by those of the vibrational peaks obtained in this study. The lowest wavenumber is theoretically estimated to be ~100 cm$^{-1}$, leading to the theoretical spectral bandwidth of ~3,000 cm$^{-1}$.

*2: The spectral bandwidth may be expanded to ~2,400 cm$^{-1}$ because the idler and signal tuning range of OPO (the laser source in Ref. 5) are 2,100-4,500 cm$^{-1}$ and 5,000-7,400 cm$^{-1}$, respectively.

**Supplementary Note 4: Retrieval of MIR interferogram**

The raw temporal MIR waveform obtained by CVS contains several components including the linear MIR interferogram. The undesired components come from the additional MIR optical fields generated when the two NIR pulses are temporally overlapped in the nonlinear crystal. The situation is similar to the DFG-type interferometric (fringe-resolved) autocorrelation[7,8]. Supplementary Figure 2**a** shows the AC part of the waveform obtained by a CVS measurement. It consists of several components such as the intensity autocorrelation, high-frequency fringes in addition to the MIR interferogram[7]. Therefore, the Fourier-transformed spectrum shows the corresponding parts as shown in Supplementary Figure 2**c**. The lowest frequency part of the spectrum near zero frequency appears due to the slowly-varying intensity autocorrelation, while the spectral components which appear around 12,500 cm$^{-1}$ and 25,000 cm$^{-1}$ come from the high frequency fringes corresponding to the fundamental and second harmonic of the NIR pulses. The MIR spectrum appears at the spectral range of around 790 – 1,800 cm$^{-1}$. Since the MIR spectrum is well separated from the other spectral components in the frequency domain, we can clearly observe the MIR spectrum without the need for correction in our demonstration.

We can extract and retrieve the MIR interferogram by eliminating the unwanted components if necessary. Since the high-frequency spectra are well-separated from the MIR spectrum, low-pass filtering removes the high-frequency fringes. The intensity autocorrelation component can be removed by subtracting a fitted intensity autocorrelation curve (derived by sech$^2$ pulses) from the raw waveform. The retrieved MIR interferogram and its spectrum are shown in Supplementary Figure 2**b** and 2**c**, respectively. This retrieval would be necessary especially when measuring lower frequency part, e.g., less than ~850 cm$^{-1}$ in our demonstration, where the MIR spectrum may overlap the low frequency spurious spectral component.

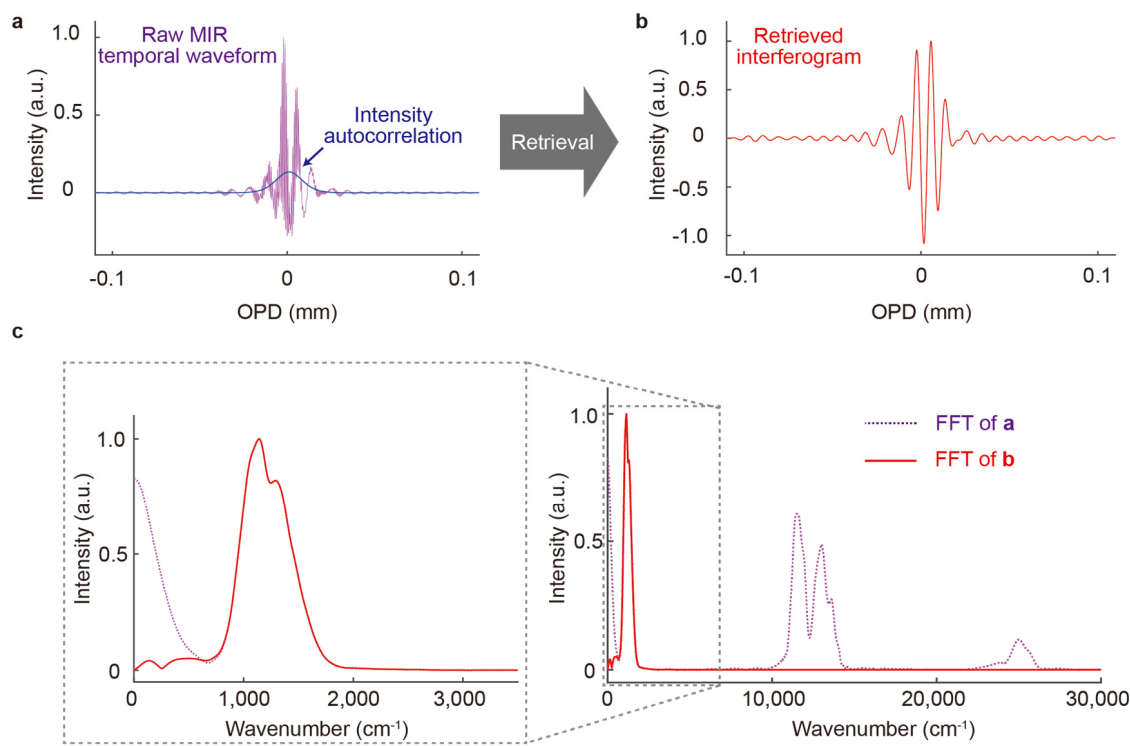

**Supplementary Figure 2|** Retrieval of MIR interferogram. **a,** AC part of the raw temporal waveform obtained by CVS. It consists of several components, which can be derived by DFG-type interferometric autocorrelation. **b,** Retrieved MIR interferogram. **c,** FFT spectra of the raw temporal waveform (dotted curve in purple) and retrieved MIR interferogram (solid curve in red).

**Supplementary References**